\documentclass[11pt,twoside]{article}

\usepackage{asp2004}
\usepackage{epsf}
\usepackage{psfig}
\usepackage{lscape}
\usepackage{graphicx}
\usepackage{natbib}

\begin{document}
\title{Evolution of rotating stars at very low metallicity}   
\author{Georges Meynet$^1$, Raphael Hirschi$^2$, Sylvia Ekstr\"om$^1$, and Andr\'e Maeder$^1$}   
\affil{$^1$Observatoire Astronomique de l'Universit\'e de Gen\`eve, CH-1290 Sauverny, Switzerland}
\affil{$^2$Dept. of Physics and Astronomy, University of Basel, CH-4056 Basel}    

\begin{abstract} 
At very low metallicity, the effects of differential rotation have a more important impact on the evolution of stars than at high metallicity. Rotational mixing leads to the production of great quantities of helium and of primary $^{14}$N by
massive stars. Rotation
induces important mass loss and allows stars to locally strongly enrich the interstellar medium in CNO elements.
Stars formed from interstellar clouds enriched by the winds of fast rotating massive stars would present
surface abundances similar to those of C-rich extremely metal-poor stars. C-rich stars can also
be formed by mass accretion in a binary system where the primary would be a fast rotating intermediate mass star in the early-AGB phase.
Fast rotation may also lead to the formation of collapsars even at very low metallicity and make the most massive stars avoid the pair instability.   

\end{abstract}


\section{Shellular rotation at very low metallicity}

The evolution of non-rotating extremely metal--poor stars has been described recently by many authors (\citeauthor{We00} \citeyear{We00}; \citeauthor{Chi02}
\citeyear{Chi02}; 
 \citeauthor{No03} \citeyear{No03}; \citeauthor*{Tu03}
\citeyear{Tu03};
\citeauthor{Pi04} \citeyear{Pi04}). Here we shall concentrate on the evolution of rotating stars at very low metallicity, a subject which, up to now, has not been so often discussed in literature
(although see \citeauthor{He02} \citeyear{He02}; \citeauthor{Ma03} \citeyear{Ma03} and the contributions
by Ekstr\"om, Woosley and Yoon in the present volume).

In the models presented in this paper,
the effects of the centrifugal acceleration in the stellar structure equations are accounted for
as explained in Kippenhahn and Thomas \citep{KippTh70}.
The equations describing the transport of the chemical species and angular momentum resulting from
meridional circulation and shear turbulence
are
given in Zahn~\citep{Za92} and Maeder \& Zahn~(\citeyear{MZ98}). The expressions for
the diffusion coefficients are taken from Talon and Zahn~\citep{TalonZ} and Maeder~\citep{Mae97}. 
The effects
of rotation on the mass loss rates is taken into account as explained in Maeder and Meynet~(\citeyear{MMVI}).

These models have been compared with stellar observations in the Galaxy and the Magellanic Clouds. They
are able to account for many observational constraints that
non--rotating models cannot fit: they can reproduce surface enrichments (\citeauthor{Hel00}
\citeyear{Hel00}; \citeauthor{MMV}
 \citeyear{MMV}), the blue to red
supergiant ratios at low metallicity (\citeauthor{MMVII} \citeyear{MMVII}), the variation with the metallicity of Wolf-Rayet populations
and of the number ratios of type Ibc to type II supernovae (\citeauthor{MMXI} \citeyear{MMXI}). Having checked that
these models compare reasonably well in the metallicity range between $Z$=0.004 and 0.020, let us now explore what
are their predictions for much lower metallicities.

Let us note before, that
if the effects of rotation are already 
quite significant at high metallicity,
one expects them to be even more important at lower metallicity. 
For instance, it was shown in previous
works that 
the chemical mixing becomes more efficient at lower
metallicity  for a given  initial mass and velocity 
(\citeauthor{MMVII} \citeyear{MMVII}; \citeauthor{MMVIII} \citeyear{MMVIII}). 
This comes from the fact that
the gradients of $\Omega$ are much steeper in the lower metallicity
models, so they trigger more efficient shear mixing.
The gradients are steeper because 
less angular momentum is transported outwards by
meridional currents, whose velocity
scales as the inverse of the density in the outer layers
(see the Gratton-\"Opick term in the expression for the meridional velocity in Maeder \& Zahn~\citeyear{MZ98}).
An interesting consequence of this greater mixing efficiency is the possibility for
metal-poor rotating stars to produce important amounts of primary nitrogen (\citeauthor{MMVIII} \citeyear{MMVIII}). 

\subsection{The case of massive stars}

\begin{figure}[!t]
\plottwo{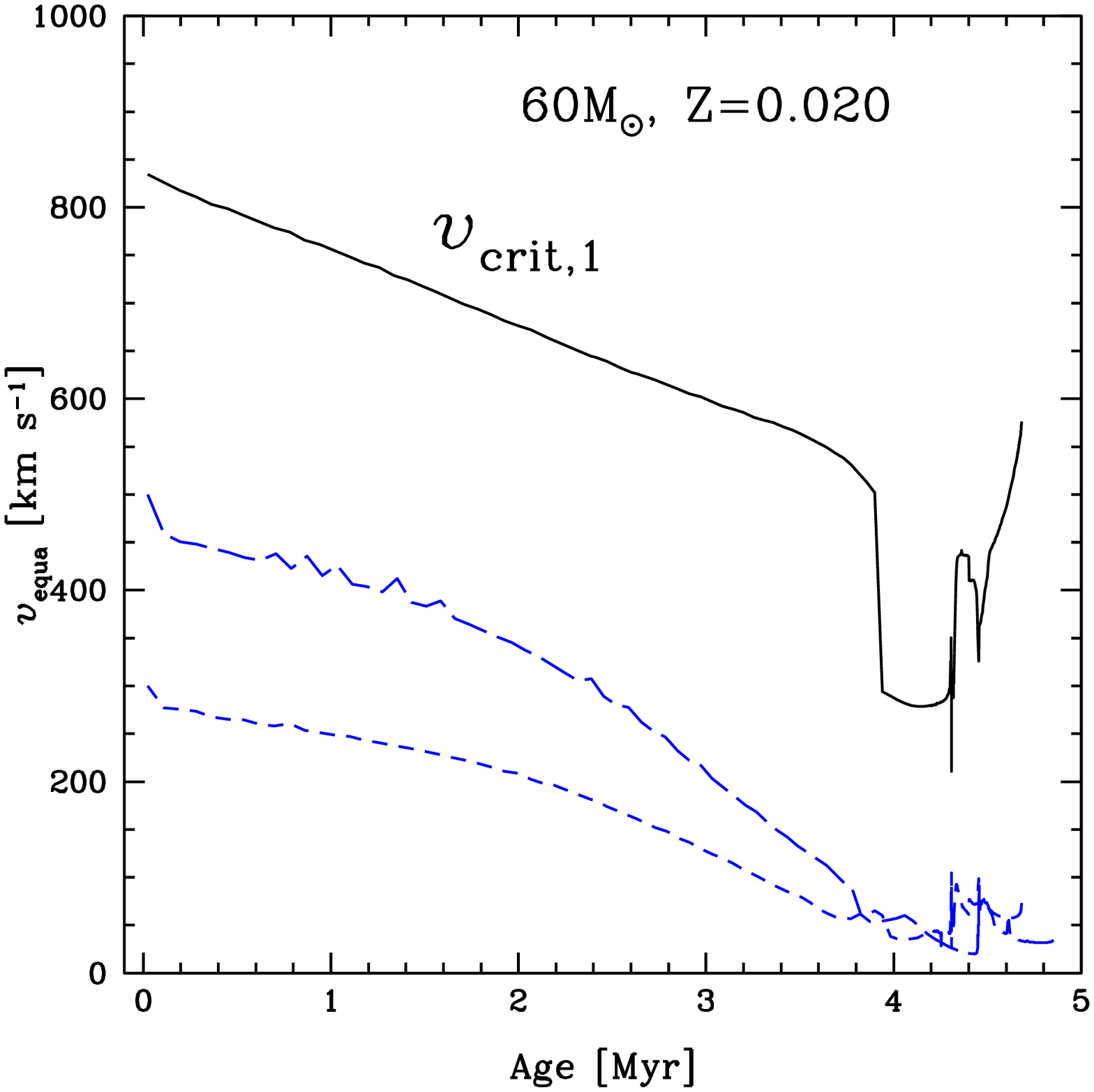}{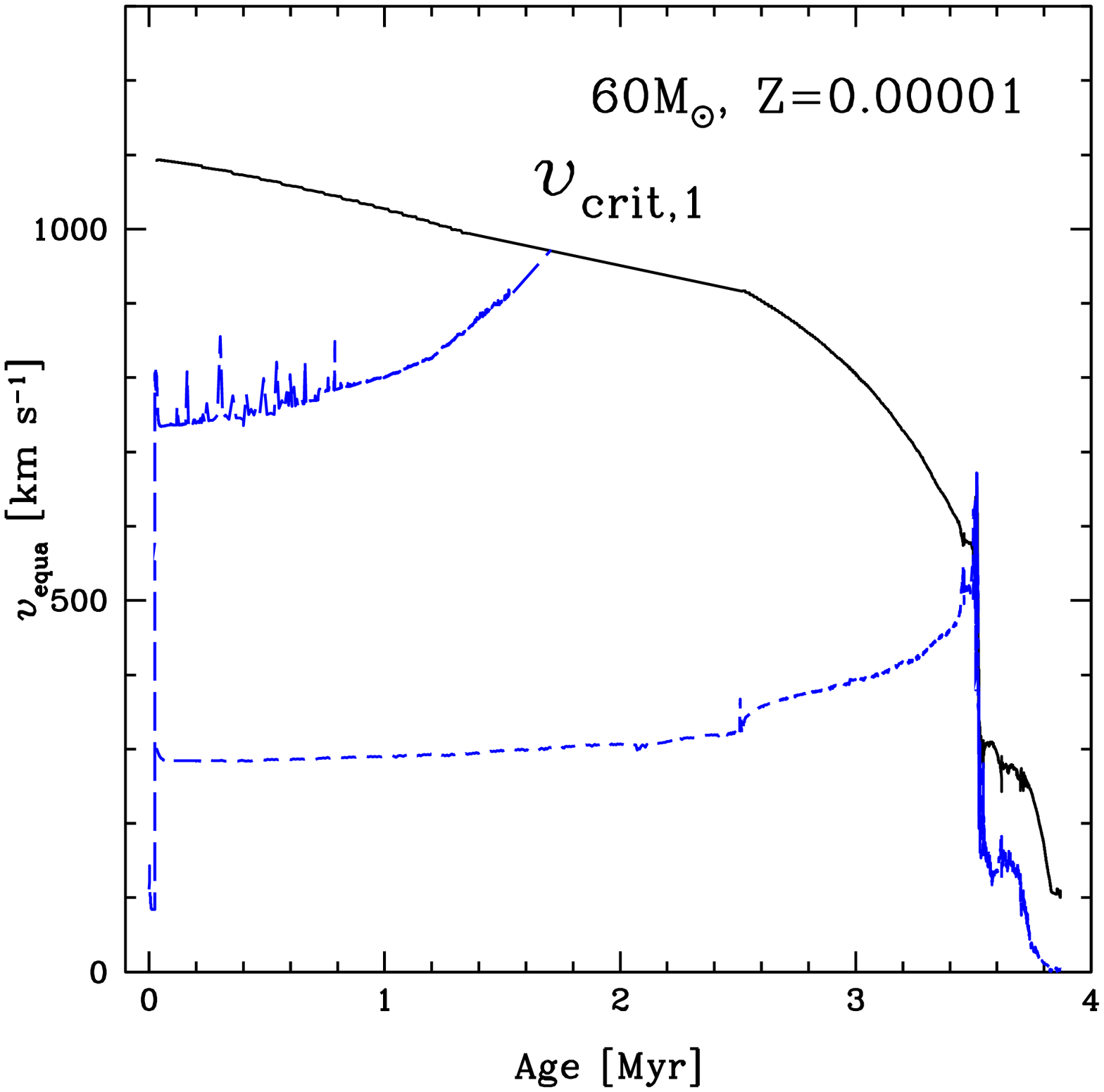}
\caption{{\it Left :} Evolution of the surface equatorial velocities at the surface of 60 M$_\odot$ models at solar metallicity with $\upsilon_{\rm ini}=$ 300 (short-dashed line) and 500 km s$^{-1}$ (long-dashed line). The continuous line shows the evolution of the equatorial critical velocity.
{\it Right :} Same as the left part of the figure, for 60 M$_\odot$ at $Z=10^{-5}$ with $\upsilon_{\rm ini}=$ 300 (short-dashed line) and 800 km s$^{-1}$ (long-dashed line).
}
\label{vcrit}
\end{figure}

\begin{figure}[!t]
\plotfiddle{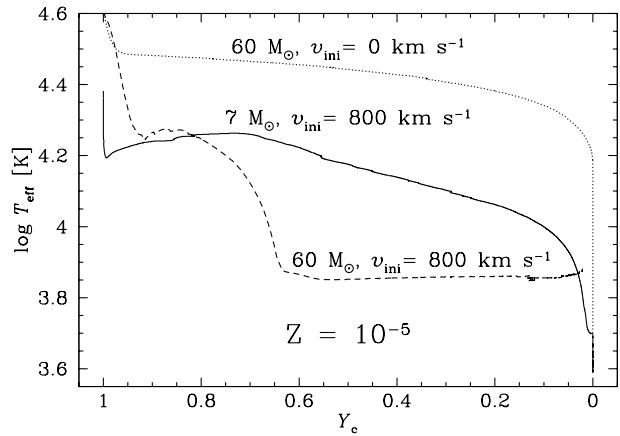}{5cm}{-90}{30}{30}{-120}{170}
\caption{Evolution of $\log T_{\rm eff}$ as a function of $Y_{\rm c}$, the mass fraction of
$^4$He at the centre, for a non-rotating (dotted line) and rotating (dashed line)
60~M$_\odot$ model at $Z=~10^{-5}$. The continuous line show the case of a 7 M$_\odot$
stellar model.}
\label{teff}
\end{figure}

In very metal-poor massive stars,  
rotation can trigger important mass loss in many ways. 
The first way, paradoxically, is linked to the fact
that very metal-poor  
stars are believed to lose little mass by radiatively driven stellar winds.
Since metal--poor stars lose little mass, 
they also lose little angular momentum (if rotating),
so they have a greater chance of
reaching the break-up limit during the Main Sequence phase. 
At break-up, the outer stellar layers become unbound and 
are ejected whatever their metallicity.  
The break-up is reached more easily when the wind anisotropy induced by rotation
are taken into account (Maeder~\citeyear{Mae99}).

In Fig.~\ref{vcrit}, the evolution of the surface velocity for 60 M$_\odot$ stellar models at two different metallicities is shown. We see that at solar metallicity, the mass loss rates are so high that, even starting with an initial velocity of 500 km s$^{-1}$, the star does not reach the critical limit. As explained above, the situation is quite different at lower $Z$, due
to the metallicity dependence of the mass loss rates: for the metallicity range between 10$^{-4}$ and 1 $\times$ Z$_\odot$, \citeauthor{Ku02} (\citeyear{Ku02})
obtains $\dot M(Z) \propto (Z/Z_\odot)^{0.5} \dot M(Z_\odot)$. 
Starting
with $\upsilon_{\rm ini}=$ 300 km s$^{-1}$, the star reaches the critical limit at the end of the Main-Sequence phase. With an initial velocity of 800 km s$^{-1}$ the critical limit is reached at a much earlier time.

\begin{figure}[!t]
\plotone{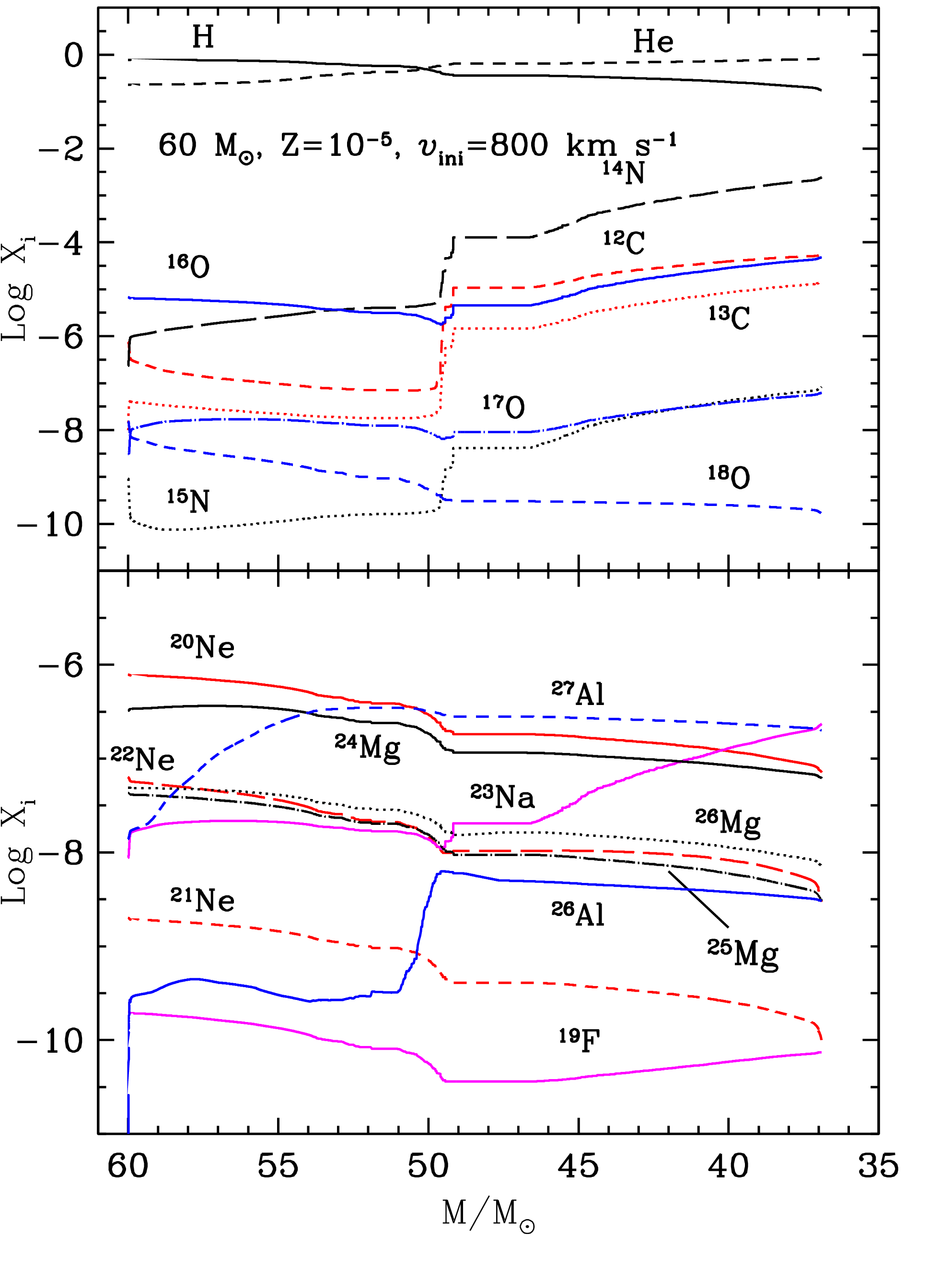}
\caption{Evolution as a function of the remaining mass of the surface abundances in mass fraction of various elements for a 60 M$_\odot$ stellar model at $Z=10^{-5}$ and with $\upsilon_{\rm ini}=800$ km s$^{-1}$.
}
\label{surabond}
\end{figure}

Another way for rotation to trigger enhancements of the mass loss
comes from the mixing induced by rotation. In general, rotational mixing
favours the evolution into the red supergiant stage (see \citeauthor{MMVII} \citeyear{MMVII}), where mass loss is higher. This is illustrated in Fig.~\ref{teff}. One sees that 
the 60 M$_\odot$ non-rotating model remains on the blue side during 
the whole core He-burning phase, while
the 800 km s$^{-1}$ model at $Z$ = 10$^{-5}$
starts its journey toward the red side of the HR diagram early in the core helium burning
stage, when $Y_{\rm c} \simeq 0.67$
($Y_{\rm c}$ is the mass fraction of helium
at the centre of the stellar model). The same is true for the corresponding model at 
$Z$ = 10$^{-8}$. Let us recall that this behaviour is linked to 
the rapid disappearance of the intermediate convective 
zone associated to the H-burning shell. 

Rotational mixing also
enhances the metallicity of the surface of the star and, in this way, may boost radiatively driven
stellar winds. Figure~\ref{surabond} shows the evolution as a function of the remaining mass of the abundances at the surface of a rotating 60 M$_\odot$ stellar model at Z=$10^{-5}$ with $\upsilon_{\rm ini}$= 800 km s$^{-1}$. During a first phase, the actual mass decreases from 60 to about 51 M$_\odot$, and the surface
metallicity (here defined as the sum of the mass fractions of all the elements except hydrogen and helium) remains equal to the initial one. The changes of the surface abundances are due to rotational mixing and
reflect the arrival at the surface of CNO processed material (decrease of $^{12}$C and
$^{16}$O abundances and increase of $^{14}$N abundance, while the sum of CNO elements remains constant).
The mass lost during this phase (about 9 M$_\odot$) results from radiatively driven stellar winds and 
evolution at the break-up limit.

When the actual mass is around 51 M$_\odot$, the star is at the middle of the core He--burning phase 
($Y_{\rm c}$ equal to 0.45) and with log $T_{\rm eff}$=3.850. From this stage on, an outer convective zone deepens in mass, dredging-up material
to the surface. This produces the sharp increase in the surface abundances in $^{12}$C,
$^{13}$C, $^{14}$N, and $^{15}$N. Then, the total amounts of heavy elements increases up to a value corresponding to more than 240 times the initial heavy element mass fraction. As can be seen from Fig.~\ref{surabond}, the wind is strongly enriched in carbon,
nitrogen, sodium and aluminium and also, to a less extent, in oxygen.

\begin{figure}[!t]
\plotone{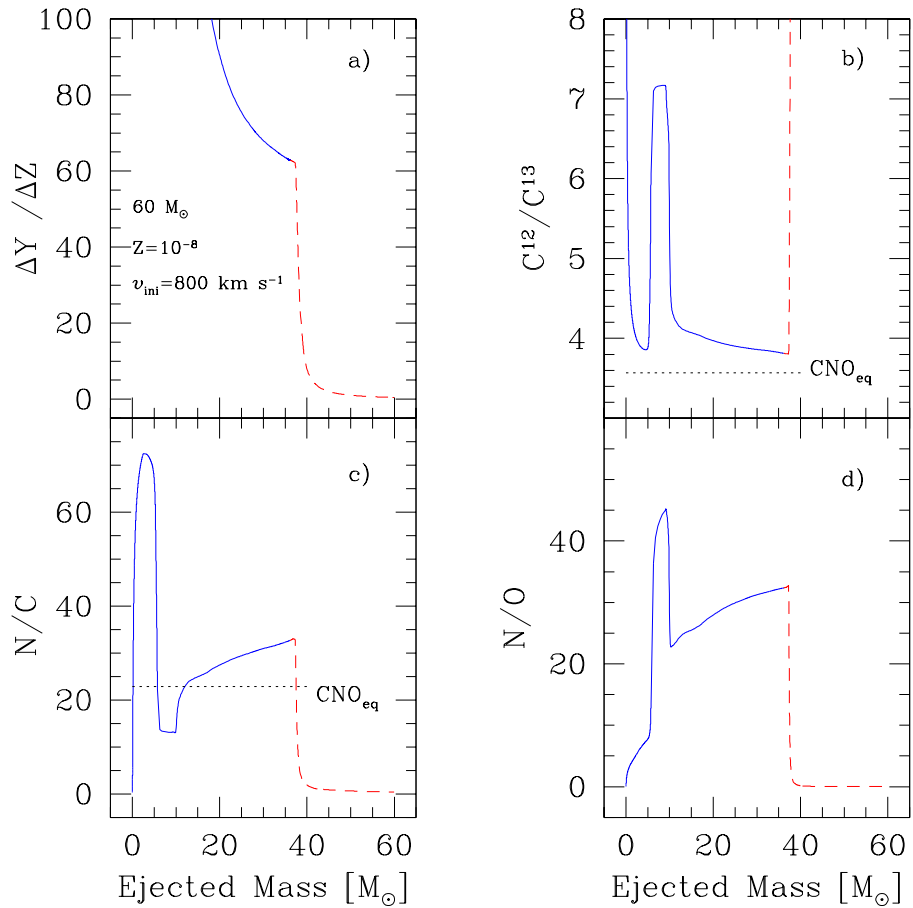}
\caption{Variation as a function of the  ejected mass of: {\bf a)} the ratio of the newly synthesized helium $\Delta Y$ to the newly synthesized heavy elements
$\Delta Z$ in the ejecta; {\bf b)} the ratio of $^{12}$C/$^{13}$C; {\bf c)} the ratio of nitrogen to carbon; 
{\bf d)} the ratio of nitrogen to oxygen. The case for
a rotating 60 M$_\odot$ model at Z=$10^{-8}$ is shown. The continuous line corresponds to the mass ejected by stellar winds, the dashed line shows how the ratios vary when
the mass from the presupernova
model is added. The dotted lines indicate the ratios obtained in the convective core of the star, when the mass fraction of hydrogen at the centre is 0.34. They correspond
to CNO equilibrium values. 
The corresponding CNO equilibrium value for the N/O ratio is 152.}
\label{aejrot}
\end{figure}

It is interesting to look at the way the ratios $\Delta Y/\Delta Z$ (mass fraction of newly synthesized helium $\Delta Y$ to the mass fraction of newly synthesized heavy elements $\Delta Z$), $^{12}$C/$^{13}$C, N/C and N/O (in mass fraction) vary as a function of the mass ejected in the wind during the stellar lifetime.
This is shown in Fig.~\ref{aejrot} for the case of the 60 M$_\odot$ stellar model at $Z=10^{-8}$ with $\upsilon_{\rm ini}=800$ km s$^{-1}$. At the beginning $\Delta Y/\Delta Z$ is not defined, it would be equal to 0/0. Then it tends to infinity when some newly synthesized helium appears at the surface, while no new heavy elements have yet reached the surface. When the ejected mass approaches about 18 M$_\odot$, $\Delta Y/\Delta Z \sim 100$, and at the end of the stellar lifetime $\Delta Y/\Delta Z \sim 65$ in the wind ejecta. The new heavy elements at this stages are in the form of new CNO elements. No newly synthesized iron is ejected in the winds. 

The initial ratios for $^{12}$C/$^{13}$C, N/C and N/O are respectively 75.5, 0.31 and 0.03. The variations of the $^{12}$C/$^{13}$C and N/C ratios are quite rapid
(see Fig.~\ref{aejrot}). Both ratios change by more than an order of magnitude in the wind ejecta, when only 6 M$_\odot$ are lost.
As expected from nuclear physics, 
at the beginning, the variation of the N/O ratio is relatively slow, then, the ratio varies
in a very short timescale by more than two orders of magnitude.

When about 6 M$_\odot$ have been lost,
the star encounters the Humphreys-Davidson limit and loses in a short while about 4-5 M$_\odot$ of its H-rich envelope. No big changes of the N/C and N/O ratios are seen, the $^{12}$C/$^{13}$C increases by a little less than a factor 2. 

The last phase, which is also the one during which the greatest amount of mass is lost, begins when about 10 M$_\odot$ have been ejected. During this phase primary nitrogen, accompanied by carbon and oxygen synthesized in the He-burning core, arrives at the surface.
On average, values of $^{12}$C/$^{13}$C equal to about 4 are obtained, while the N/C and N/O ratios have values of about 30.

The dashed line in Fig.~\ref{aejrot} shows how the addition of material from the presupernova model changes the ratios. When the supernova ejecta are taken into account, much lower values of $\Delta Y/\Delta Z$, N/C and N/O are obtained, and much higher
$^{12}$C/$^{13}$C ratios.
The addition of a very small amount of 
mass ejected by the supernova already changes a lot the ratios. For instance, the addition of only 4 M$_\odot$ ejected at the time of the supernova explosion would lower the N/C ratio from the value of 33 to about 2.

\begin{figure}[!t]
\plottwo{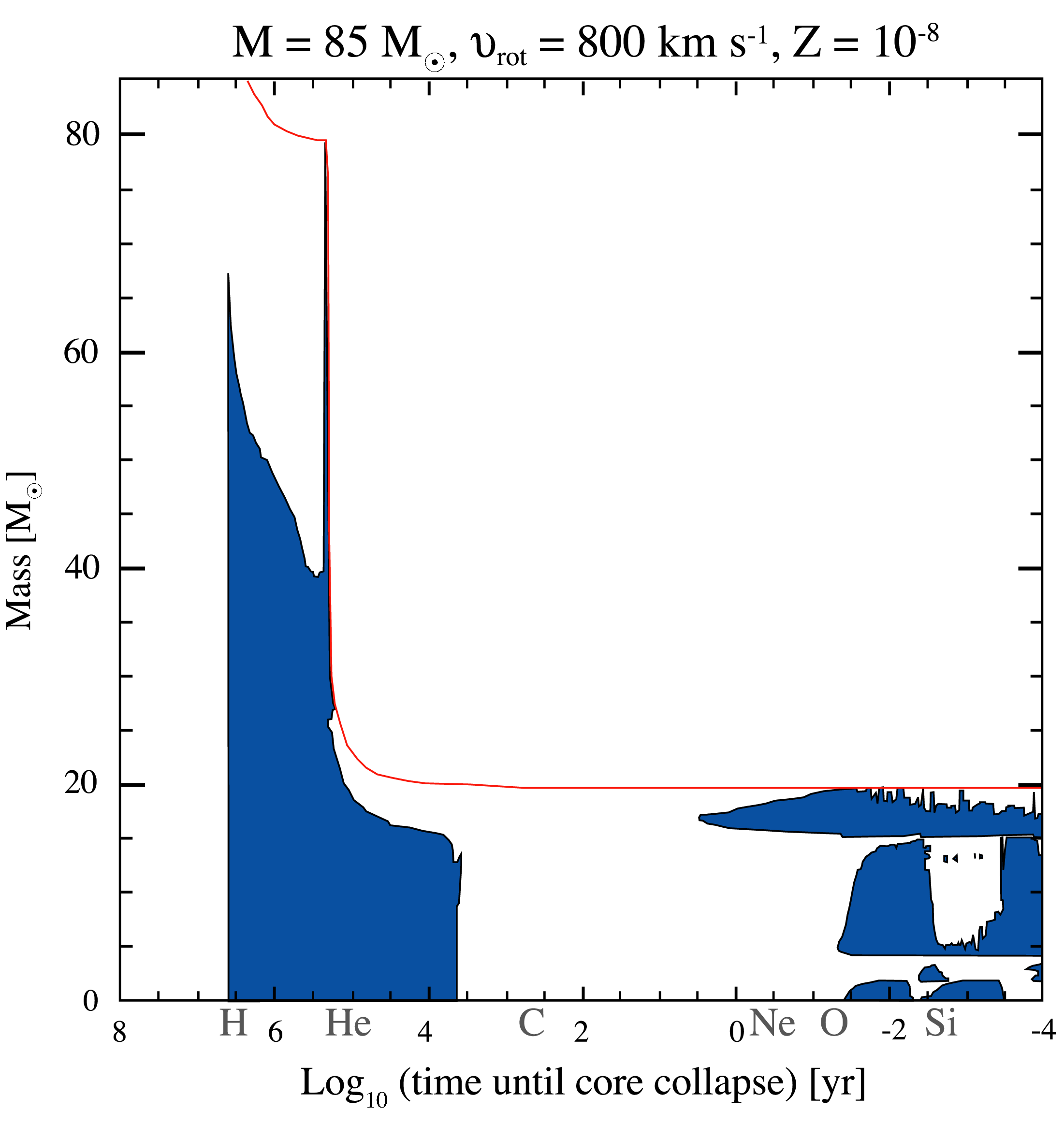}{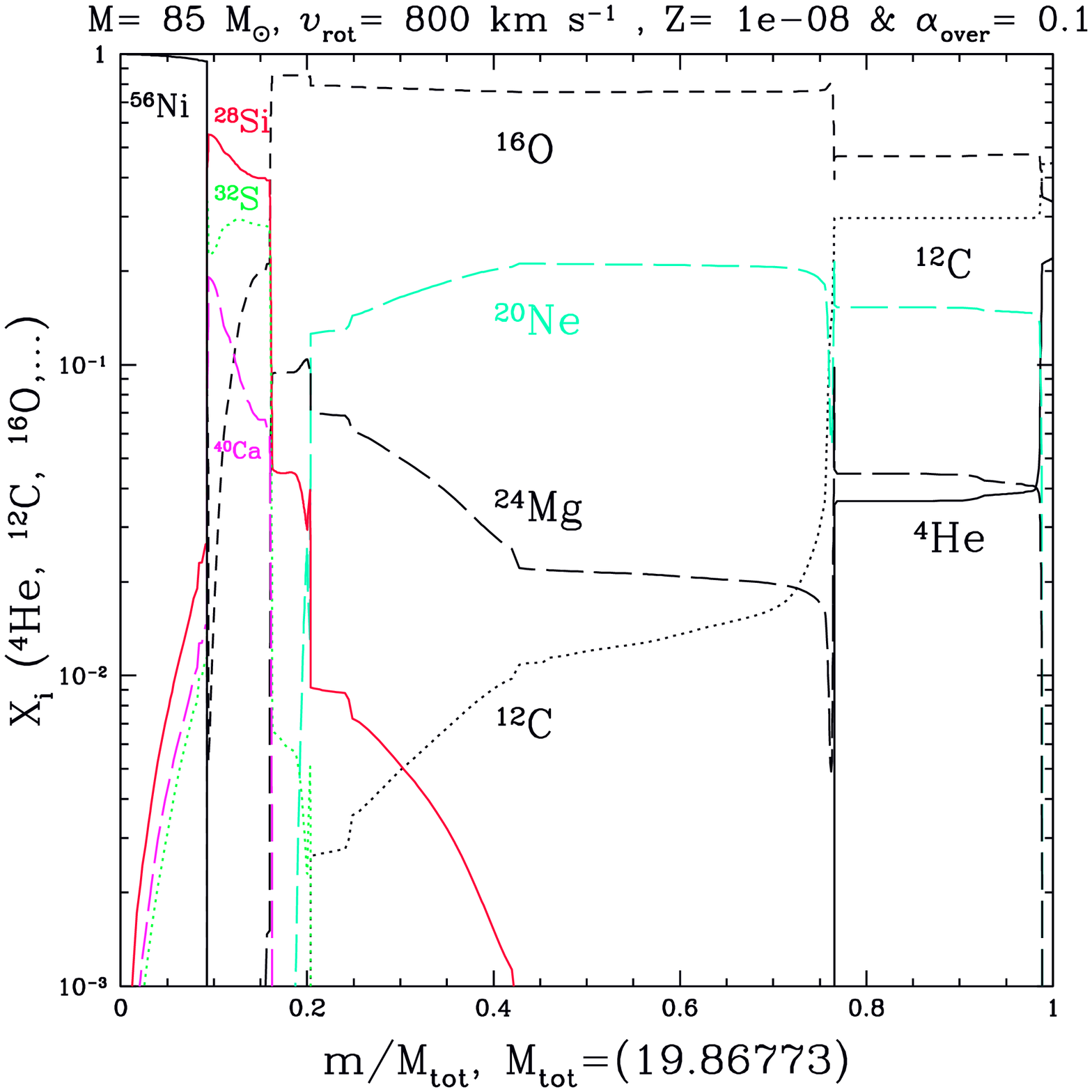}
\caption{{\it Left :} Evolution of the total mass for a rotating 85 M$_\odot$ stellar model as a function of the logarithm of the time before the supernova explosion. The dark areas correspond to convective zones. {\it Right :} Variation of the chemical composition as a function of the lagrangian mass normalised to the total actual mass at the end of the core Si--burning stage for the same model as in the left part of the figure. Both figures are from Hirshi (2005).
}
\label{m85}
\end{figure}

The importance of the various enrichments depends on metallicity, the initial mass and
the initial velocity considered. The case of the 85 M$_\odot$ models
at $Z=10^{-8}$, with $\upsilon_{\rm ini}$ = 800 km s$^{-1}$ is shown in Fig.~\ref{m85} (Hirschi~\citeyear{HIP}).
This model loses more than 75\% of its initial mass, and will terminate its life as a WO star. Interestingly,
its angular momentum content at the presupernova stage is sufficient for giving birth to a collapsar 
(see also \citeauthor{HMMXIII} \citeyear{HMMXIII}) now considered as among the best candidates for the progenitors of Gamma Ray Bursts (see Woosley~\citeyear{Wo93}; Mazzali et al. 2003). Therefore, according to such models (without magnetic fields), GRB may be produced even
at very low metallicity.

\subsection{Do massive very metal--poor, stars end their lives as pair--insta\-bility supernovae?}

Might the important mass loss undergone by rotating models prevent 
the most massive stars from going through pair instability?
According to Heger \& Woosley (\citeyear{He02}), progenitors of pair--instability supernovae
have helium core masses
between $\sim$64 and 133 M$_\odot$. This corresponds to initial masses between about 140 and 260 M$_\odot$.
Thus the question is whether
stars with initial masses above 140 M$_\odot$ can lose a sufficient amount of mass
to have a helium core that is less than about 64 M$_\odot$
at the end of the core He-burning phase. 
From the values quoted above, it would imply the loss of more than
(140-64) M$_\odot$=76 M$_\odot$, which represents about 54\%
of the initial stellar mass. 
From the computations presented here, where
a 85 M$_\odot$ loses more than 75\% of its initial mass,
one can expect that such a scenario is possible. However, 
more extensive computations are needed to
check in which metallicity range and for which initial velocities
rotational mass loss could indeed prevent the most massive stars
from going through pair instability. Let us note that, for the same initial velocities as
considered here, Pop III stellar models would not avoid the pair instability (see the
contribution by Ekstr\"om et al. in this volume), however the results might be different for higher initial velocities.

Would pair instability be avoided, this would explain why
the nucleosynthetic signature of pair--instability supernovae is not observed
in the abundance pattern of the most metal--poor halo stars known up to now.
At least this rotational mass loss could restrain the mass range for the progenitors
of pair--instability supernovae, pushing the minimum initial mass needed for such a scenario to occur to higher values. Moreover,
when the initial mass increases,
the luminosity of the star comes
nearer to the Eddington limit. When rotating, such stars will then encounter
the $\Omega\Gamma$-limit (\citeauthor{MMVI} \citeyear{MMVI}) and very likely undergo strong mass losses.

\subsection{The intermediate mass stars}

\begin{figure}[!t]
\plottwo{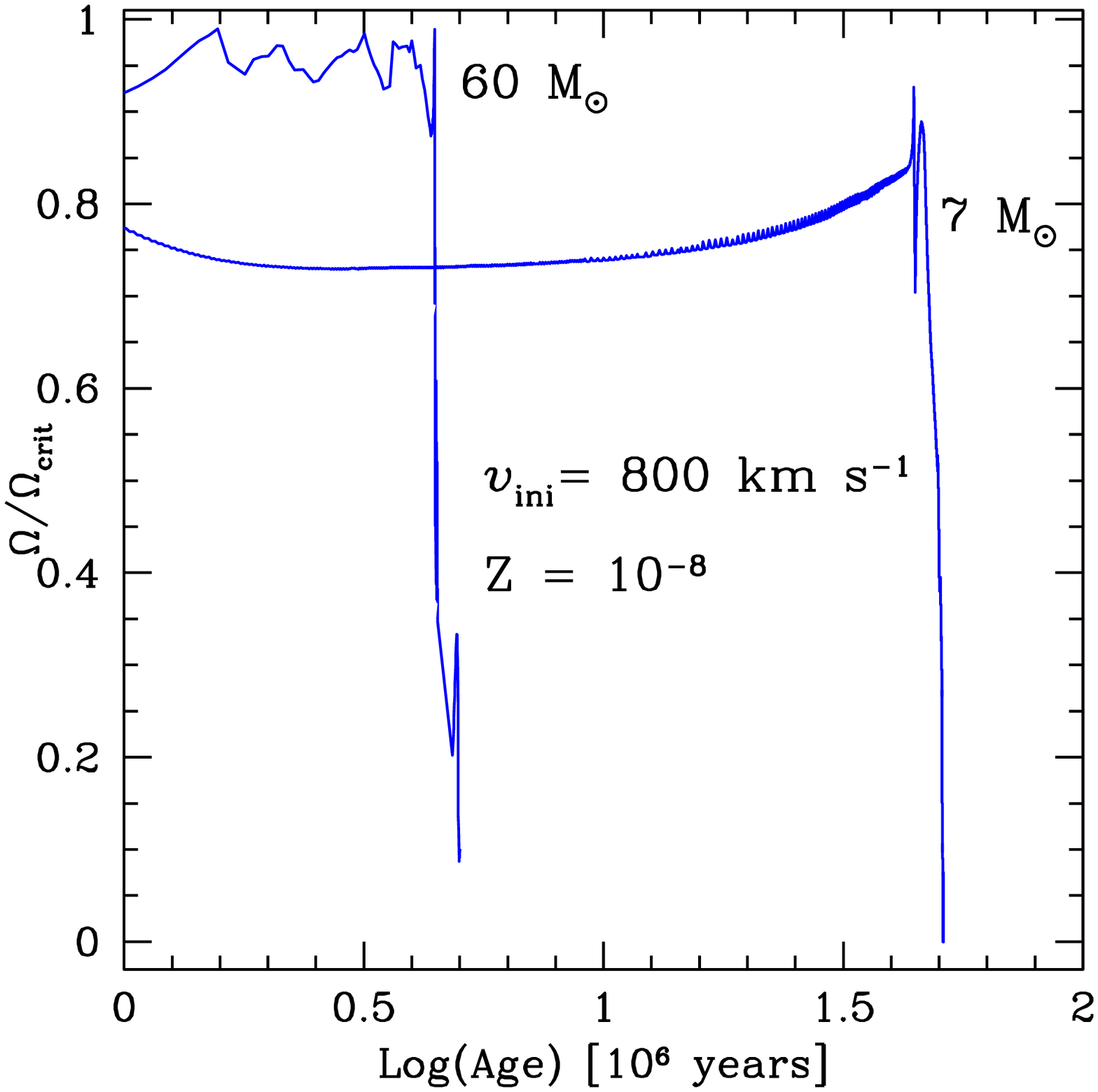}{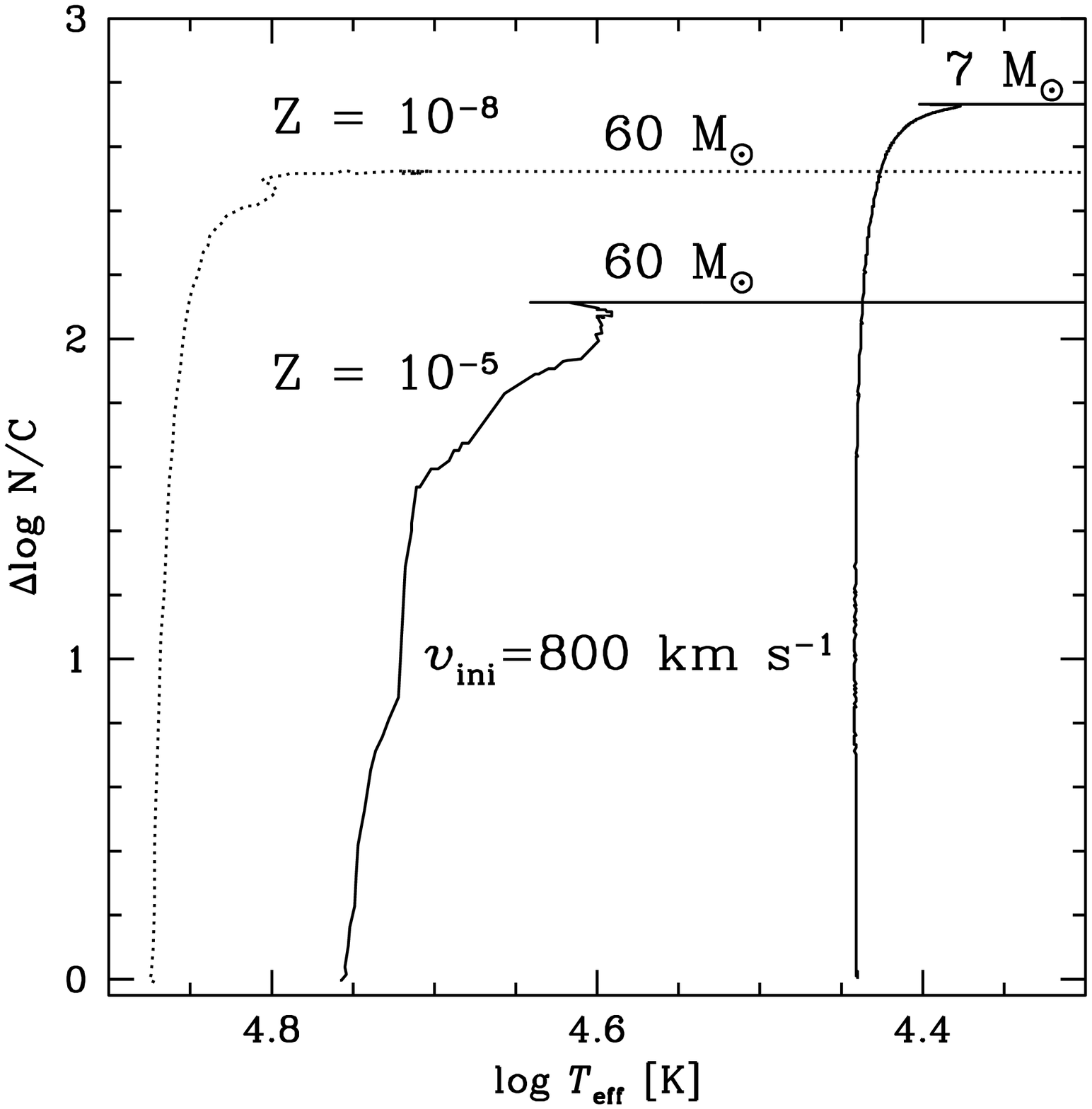}
\caption{{\it Left:} Evolutions of the ratio of the actual angular velocity to the critical angular velocity at the surface of a rotating 60 and 7 M$_\odot$ stellar model at very low metallicity.{\it Right:} Evolution 
as a function of the effective temperature
of the N/C ratio ($\Delta {\rm log} {\rm N/C}={\rm log (N/C)}-{\rm log (N/C)}_{\rm ini}$) at the surface of the two same models. The dotted line corresponds to a 60 M$_\odot$ stellar model at $Z=10^{-8}$ with the same initial velocity. 
}
\label{sept60}
\end{figure}

Rotation not only affects the evolution of massive stars but also that of intermediate mass stars.
In Fig.~\ref{sept60}, the evolution of the equatorial velocity and of the nitrogen to carbon ratio at the surface
of a 7 M$_\odot$ stellar model is shown. They are compared to the evolution of the corresponding 
quantities in 60 M$_\odot$ models.
While the 60 M$_\odot$ model rapidly reaches the break-up limit,
the 7 M$_\odot$ maintains during most of the Main-Sequence phase a value of $\Omega/\Omega_{\rm crit}$ equal to 0.8 at the surface. Let us note that the outer layers of the 7 M$_\odot$ stellar model are denser than in more massive stellar models. Thus, the Gratton-\"Opick term ($\propto 1/\rho$) in the expression of the meridional velocity is therefore smaller, making the outwards transport of the angular momentum less efficient. In the 7 M$_\odot$ stellar model, 
the gradients of $\Omega$ are steeper
and the chemical mixing is thus more efficient. This can be seen in the right part of Fig.~\ref{sept60}.

\section{Links with the C-rich extremely metal--poor stars}

If most stars at a given [Fe/H] present great
homogeneity in composition, a small group, comprising about 20 - 25\% of the stars
with [Fe/H] below -2.5, show very large enrichments in carbon.
These stars are known as C-rich
extremely metal--poor (CEMP) stars. The observed [C/Fe] ratios
are between
$\sim$2 and 4, showing a large scatter.
Other elements, such as nitrogen and oxygen (at least in the few cases
where the abundance of this element could be measured), are also highly  
enhanced. Interestingly, 
the two most metal--poor stars known up to now, the Christlieb star
(HE 0107-5240), a halo giant with [Fe/H]=-5.3 (Christlieb et al.~2004),
and the subgiant or main-sequence star HE 1327-2326 with [Fe/H]=-5.4 
(\citeauthor{Fr05} \citeyear{Fr05})
belong to this category.

A few CEMP stars are Main-Sequence or subgiant stars like the Frebel star.
At these evolutionary stages,
no process occurring in the star itself can explain the surface abundances.
For these stars at least,
the surface abundances reflect the abundances of the protostellar cloud from which the star formed, or
the surface abundances of a more evolved companion whose part of the mass has been accreted by the CEMP star. 

To explain the very high C-abundance, various models have been proposed. 
For instance, Umeda \& Nomoto~(2003) propose that the cloud from which HE 0107-5240
formed was enriched by the ejecta of one Pop III 25 M$_\odot$ star, which had exploded with low
explosion energy (on the order of 3 $\times 10^{50}$ erg) and experienced 
strong mixing and fallback at the time of the supernova explosion.
This would mimic asymmetric explosions. Limongi \& Chieffi~(2003)
suggest that the cloud was enriched by the ejecta of two Pop III supernovae
from progenitors
with masses of about 15 and 35 M$_\odot$. 
Some authors have proposed that this particular 
abundance pattern results from accretion of 
interstellar material and from a companion (for instance an AGB star, as proposed by \citeauthor{Su04} \citeyear{Su04}).

\begin{figure}[!t]
\plottwo{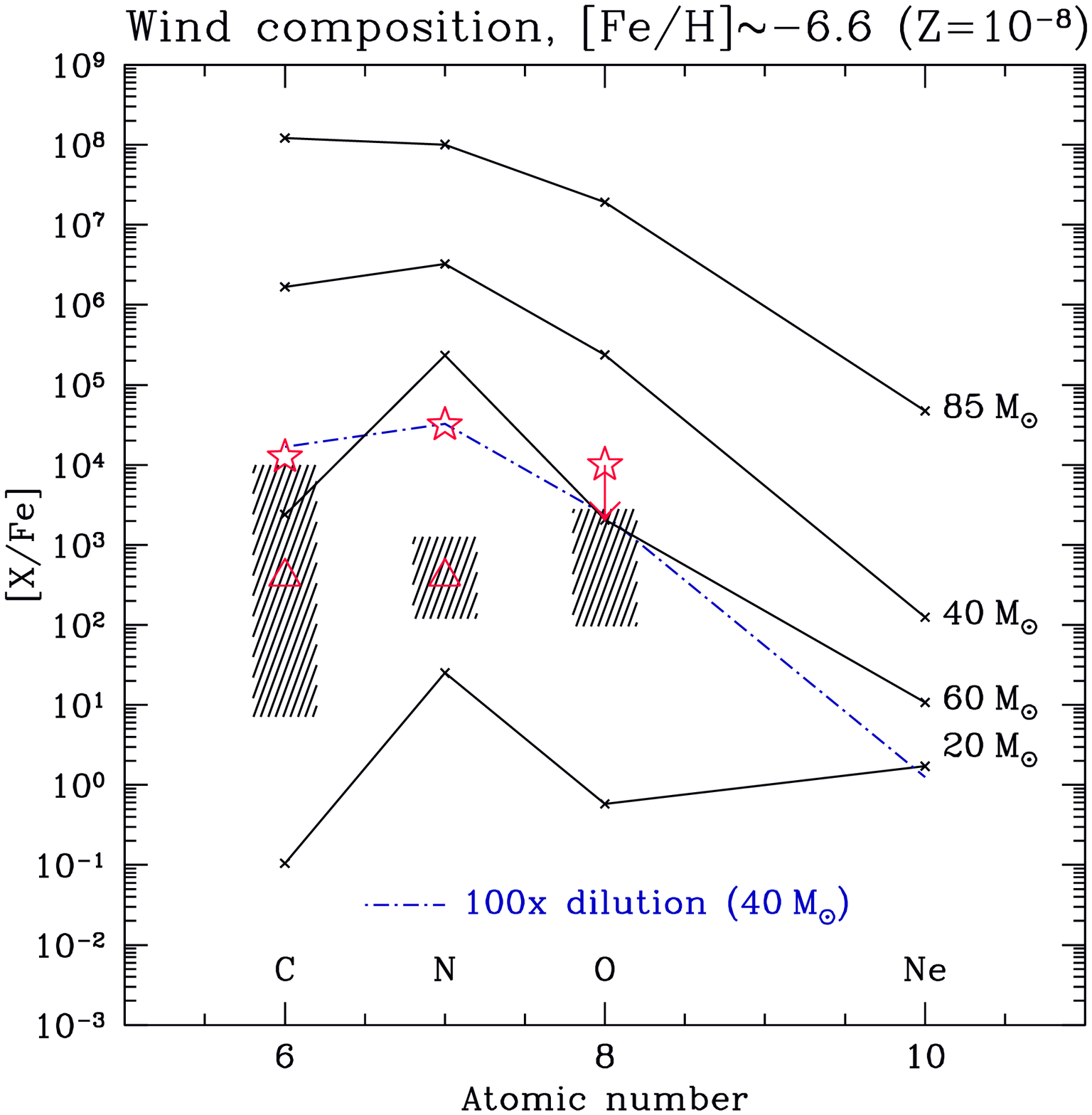}{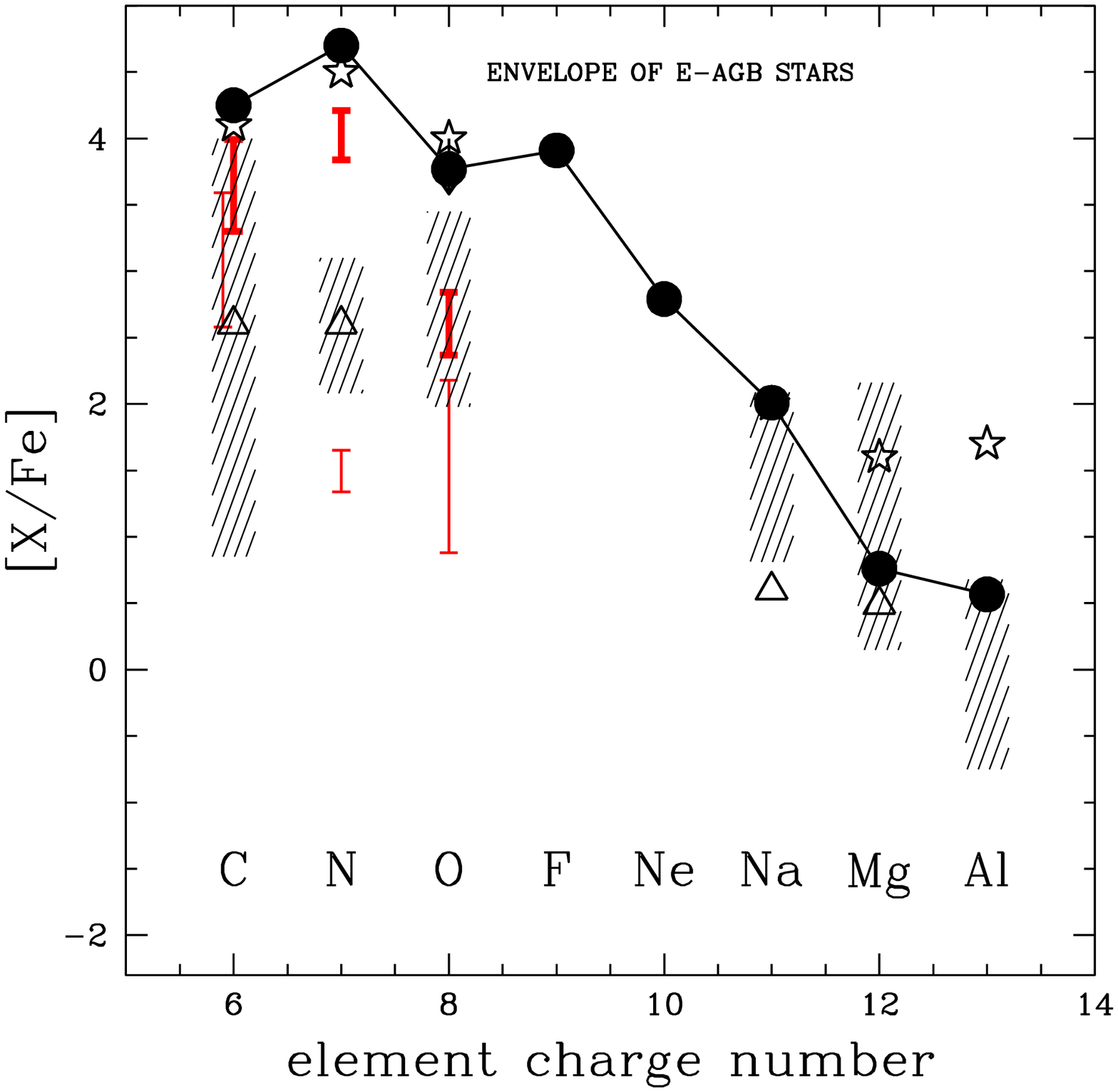}
\caption{{\it Left :}  
Chemical composition of the wind
of rotating models (continuous lines) of various initial masses from the work of Hirschi (2005).
When the wind material of the 40 M$_\odot$ stellar model is mixed with hundred times more of interstellar material, the dot-dashed curve is obtained.
The hatched areas correspond to the range of values
measured at the surface of giant CEMP stars: HE 0107-5240, [Fe/H]$\simeq$~-5.3 
\citep{christ04};
CS 22949-037, [Fe/H]$\simeq$~-4.0 (\citeauthor{norr01} \citeyear{norr01}; \citeauthor{dep02} \citeyear{dep02}); CS 29498-043, [Fe/H]$\simeq$~-3.5 \citep{aoki04}. The empty triangles (\citeauthor{Pl05} \citeyear{Pl05}, [Fe/H]$\simeq -4.0$) 
and stars (\citeauthor{Fr05} \citeyear{Fr05},  [Fe/H]$\simeq -5.4$, only an upper limit is given for [O/Fe]) correspond to
non-evolved CEMP stars (see text).{\it Right :}
Chemical composition of the envelopes of E-AGB stars compared to abundances
observed at the surface of CEMP stars (references as in the left
part of the figure). The continuous line shows the case
of a 7 M$_\odot$
at $Z=10^{-5}$ ([Fe/H]=-3.6) with $\upsilon_{\rm ini}=$ 800 km s$^{-1}$.
The vertical lines (shown as ``error bars'')
indicate the 
ranges of values for CNO elements
in the stellar models of Meynet \& Maeder (2002)
(models with initial masses between 2 and 7 M$_\odot$ at $Z=10^{-5}$).
The thick and thin ``error bars'' correspond to rotating ($\upsilon_{\rm ini}$ = 300 km s$^{-1}$)
and non-rotating models.}
\label{cemp}
\end{figure}

Recently, we have explored the possibility that such CEMP stars could be formed from the wind ejecta of
massive rotating stars (Meynet et al. \citeyear{MEM05}). Let us suppose that these massive stars produce, at the end of their nuclear lifetime, a black hole, that swallows the whole final mass. In that case, the massive star
would contribute to the local chemical enrichment of the interstellar medium only through
its winds. In Fig.~\ref{cemp} the abundances in the winds of various initial mass models are
plotted (Hirschi 2005). We see that for all masses superior to about 40 M$_\odot$ enhancements of CNO elements with respect to iron are obtained. For comparison the enhancements of these elements observed at the surface of CEMP stars are indicated by hatched areas. The continuous line corresponds to the abundance in pure wind material. When the wind material of the 40 M$_\odot$ stellar model is mixed with hundred times more of interstellar material, the dot-dashed curve is obtained. We see that wind material, diluted with some interstellar matter, could reproduce the CNO abundance pattern observed at the surface of CEMP stars. 

What does happen now if the supernova also contributes to the enrichment of the cloud ?  From the mass of CO-core $M_{\rm CO}$ it is possible to infer the mass of the remnant $M_{\rm rem}$ using the relation of Arnett~(1991)
between $M_{\rm CO}$ and $M_{\rm rem}$. The value $M_{\rm rem}$ obtained is equal to 5.56 M$_\odot$
for the 60 M$_\odot$ stellar model shown in Fig.~\ref{aejrot}. In
Fig.~\ref{aejrot}, this corresponds to a total mass ejected of 54.44 M$_\odot$ (60-5.56). The values of $\Delta Y/\Delta Z$,  $^{12}$C/$^{13}$C, N/C and N/O in the ejecta are then respectively 0.68, 309, 0.5 and 0.02. 
These values are not
compatible with the ratios observed at
the surface of CS 22949-037 (Depagne et al.~\citeyear{dep02}): $^{12}$C/$^{13}$C $\sim$4,
N/C $\sim$ 3 and N/O $\sim$0.2. To obtain
a better fit in the frame of such models, much less mass from the presupernova
should be ejected, confirming the idea of strong fallback proposed by Umeda \& Nomoto~(2003).

The physical
conditions encountered in the advanced phases of an intermediate mass star
are not so different from the one realised in massive stars. Thus the same
nuclear reaction chains can occur and lead to similar nucleosynthetic products.
Also the lifetimes of massive stars (on the order of a few million years) are
not very different from the lifetimes of the most massive intermediate mass stars;
typically a 7 M$_\odot$ has a lifetime on the order of 40 Myr, only an order of magnitude higher
than a 60 M$_\odot$ star.
Moreover, the observation of s-process element overabundances at the surface of some
CEMP stars also point toward a possible Asymptotic Giant Branch (AGB) star origin.

To explore a scenario where the particular surface abundances of the CEMP star results from
a mass transfer from a primary in the AGB stage, Meynet et al.~(2005)
computed a
7 M$_\odot$ with $\upsilon_{\rm ini}=800$ km s$^{-1}$ at $Z=10^{-5}$. 
In contrast to the 60 M$_\odot$ models, the 7 M$_\odot$ stellar model 
loses little mass during the core H- and He--burning phase, so that 
the star has still nearly its whole
original mass
at the
early asymptotic giant branch stage (the actual mass at this stage is 6.988 M$_\odot$). This is because
the star never reaches the break-up limit during the MS phase (see Fig.~\ref{sept60}). Also, 
due to rotational mixing and dredge-up,
the metallicity
enhancement at the surface  
only occurs very late, when the star
evolves toward the red part of the HR diagram
after the end
of the core He-burning phase. At this stage, the outer
envelope of the star is enriched in primary CNO elements, and the surface metallicity
reaches about 1000 times the initial metallicity.
If such a star is in a close
binary system, there is a good chance that mass transfer occurs during this
phase of expansion of the outer layers. In that case, the secondary may accrete
part of the envelope of the E-AGB star. 

From the 7 M$_\odot$ stellar model, we can estimate the chemical composition 
of the envelope 
at the beginning of the thermal pulse AGB phase. Here we call
envelope all the material above the CO-core.
The result is shown in Fig.~\ref{cemp} (continuous line with solid circles).
We also plotted the values obtained from the models
of Meynet \& Maeder~(2002) for 
initial masses between 2 and 7 M$_\odot$ at $Z=10^{-5}$
and with $\upsilon_{\rm ini}=0$ and $300$ km s$^{-1}$.

We see that the envelopes of AGB stellar models with rotation 
show a chemical composition  very similar to the one observed at the surface
of CEMP stars. It is, however, still difficult to say that rotating intermediate mass star models
are better than rotating massive star models in this respect. Probably,
some CEMP stars are formed from massive star ejecta and others
from AGB star envelopes. Interestingly at this stage,
some possible ways to distinguish between massive star wind material
and AGB envelopes do appear. Indeed, we just saw above that
massive star wind material is characterized by a very low $^{12}$C/$^{13}$C ratio,
while intermediate mass stars seem to present higher values for this ratio.
The AGB envelopes would also present very high overabundances
of $^{17}$O, $^{18}$O, $^{19}$F, and $^{22}$Ne, while wind of massive rotating
stars present a weaker overabundance of $^{17}$O and depletion of 
$^{18}$O and $^{19}$F (Meynet et al.~2005).
As discussed in Frebel et al. (2005), the ratio of heavy elements, such
as the strontium--to--barium ratio, can also give clues to the origin of the material
from which the star formed. In the case of HE 1327-2326, Frebel et al.~(2005)
give a lower limit of [Sr/Ba] $> -0.4$, which suggests that strontium was not
produced in the main s-process occurring in AGB stars, thus leaving
the massive star hypothesis as the best option, in agreement with the result
from $^{12}$C/$^{13}$C in G77-61 (Plez \& Cohen 2005) and
CS 22949-037 (Depagne et al. 2002).

\section{Helium enrichment by the first stellar generations}

The question of the enrichment in newly synthesized helium by the first stellar generations was
already asked long ago by
Hoyle \& Tayler (1964).
The recent finding of a double sequence in the globular cluster $\omega$ Centauri by Anderson (1997)
(see also \citeauthor{Bedin04} \citeyear{Bedin04}) and the  further interpretation of the bluer sequence by a
strong excess of helium requires some reexamination of this very interesting question.
The bluer sequence, with a metallicity
[Fe/H]= -1.2  or $Z =  2 \cdot 10^{-3}$, implies a helium content $Y=0.38$, 
i.e. a helium enrichment $\Delta Y = 0.14$, which in turn demands a relative helium 
to metal enrichment  $\Delta Y/\Delta Z$ of the order of  70
(\citeauthor{Piotto05} \citeyear{Piotto05}). 
This value of  $\Delta Y/\Delta Z$ is enormous and more than one order of magnitude larger than the current value of $\Delta Y/\Delta Z= 4-5$ (Pagel 1992) obtained from observations 
 of extragalactic  HII regions. This low value of $\Delta Y/\Delta Z$ is  also consistent with the chemical yields from supernovae
 forming  black holes above about 20--25 M$_{\odot}$ (\citeauthor{Ma92} \citeyear{Ma92}).
 
Fast rotating massive stars at very low metallicity eject important amounts of material
rich in newly synthesized helium. As a numerical example, for the rotating 60 M$_\odot$ at $Z=10^{-8}$, 22\% of the initial mass is ejected in the form 
of newly synthesised helium by stellar winds. From Fig.~\ref{aejrot}, one also sees that the value of $\Delta Y/\Delta Z$ is equal to about 65 in the wind material. 
This shows that this type of model do appear as very promiseful for reexamining the question
of the helium enrichment by massive stars at very low Z.



\end{document}